\documentclass[12pt]{article}
\usepackage{amsmath, amssymb, graphicx, booktabs, hyperref}
\usepackage[left=3cm, right=3cm, top=2.5cm, bottom=2.5cm]{geometry}
\usepackage{subcaption}

\title{Feedback Linearization for Replicator Dynamics: A Control Framework for Evolutionary Game Convergence}

\author{Adil Faisal}
\date{August 17, 2025}

\begin{document}
\maketitle

\section*{Abstract}
This paper demonstrates the first application of feedback linearization to replicator dynamics, driving the evolution of non-convergent evolutionary games to systems with guaranteed global asymptotic stability. Replicator dynamics, while a cornerstone of evolutionary game theory, possess neutral stability at Nash equilibria \cite{fryer2018}, which causes the evolutionary process to oscillate without converging to an optimal strategy. We build a control-theoretic framework that cancels the nonlinear components in replicator dynamics, and then apply a linear feedback component to force a strategy change at the Nash equilibrium. Through Lyapunov analysis, we show global convergence from any initial conditions in the probability simplex. We illustrate this approach with a numerical example of a penalty shootout game, where we illustrate that our method guides strategies quickly to mixed Nash equilibria, while the uncontrolled dynamics oscillate. Our work serves as one of the first known connections between nonlinear control theory and evolutionary game dynamics with applications in multi-agent systems, algorithmic trading, and strategic optimization. \\

\textbf{Keywords}: Game theory, Lyapunov stability, Replicator dynamics, Nash equilibrium, Feedback linearization, Nonlinear control, Evolutionary games

\section{Introduction}
Replicator dynamics are a foundational model within evolutionary game theory that describes how frequencies of strategies change according to their fitness in a competitive context \cite{allesina2025}. While replicator dynamics are often successful at modeling strategy evolution in biological and economic systems, replicator dynamics also have a fundamental flaw: Nash equilibria are typically neutrally stable \cite{fryer2018}. Thus, strategies oscillate about optimal strategies rather than converging. More specifically, the oscillations prevent agents from converging towards optimal play and instead lead to unpredictable long-term dynamics in strategic interaction.

This paper addresses the problem of convergence by applying full state feedback linearization to replicator dynamics, signifying the first time this nonlinear control method has entered the evolution of games. This approach turns the non-convergent dynamics of replicator dynamics into a system that achieves global asymptotic stability, allowing agents to reliably converge to Nash equilibrium strategies from any initial condition in the replicator dynamics. Our method cancels the nonlinear terms responsible for oscillatory dynamics and applies linear feedback control to essentially steer the strategies towards the Nash equilibrium. With the use of Lyapunov theory, we analytically show that the proposed control law will yield global asymptotic stability to a point in the probability simplex. With this controlling law, we can surely guarantee all agents will converge to the Nash equilibrium in a game setting, where standard replicator dynamics cannot guarantee convergence. 

We illustrate our method with a penalty shootout between a striker and a goalie, using simulations to demonstrate the difference in convergence to equilibrium between this method and the uncontrolled replicator dynamics with oscillations. This work represents a new direction at the intersection of control theory and evolutionary games, with broader implications of multi-agent learning, algorithmic trading, allocation of resources in networks, and any strategic setting that attempts to achieve reliable and predictable convergence to equilibrium using evolutionary-type dynamics.

\section{Game setup and background}

The game between the players is set out to be a zero-sum game between the players. This means that the total sum of benefits of the game is zero, meaning that if one player wins, the other player must lose. In the context of a penalty shootout, this assumption works since if the striker scores, the goalkeeper loses, and vice versa. The game is also simultaneous since both the goalkeeper and the players must decide simultaneously \cite{hayes2024}. Each striker and goalkeeper has three directions to shoot the ball or dive, left, right, and center, and therefore, the game is a mixed strategy game. Since the game is a mixed strategy game, replicator dynamics can be utilized to analyze the strategies of the agents as they evolve.

Nash equilibrium (NE) is defined as an outcome when achieved, means that no player can increase their payoffs by changing their decisions unilaterally \cite{hayes2024}. In the context of this penalty shootout, since the game is noncooperative, this implies that the game has at least one point that is NE for both the striker and the goalkeeper \cite{fryer2018}.

\section{Game Dynamics}

\[
A=
\begin{bmatrix}
0 & 0.8 & 0.7 \\
0.9 & 0 & 0.2 \\
0.75 & 0.45 & 0
\end{bmatrix}
\tag{1}
\]

A is the payoff matrix defined for the striker. To account for the stochastic nature of penalty kicks, all the payoff values are set to be in the interval $[0,1]$ so that each payoff value represents the chance of a goal for a given shot direction and diving direction \cite{monstad2023}. If the player shoots to the left and the goalkeeper stays in the center, the payoff for the striker is 0.8.

This allows for misses or deflections in shots to be accounted for during the penalty, since each value in matrix A is the chance of scoring a goal given any mix of decisions made by the players. A key assumption in the game is that if the striker's shot direction and the goalkeeper's diving direction are the same, the payoff is zero, so neither player wins nor loses.

The goalkeeper's playoff matrix B is defined by equation (2) since the game is asymmetric, meaning that the roles cannot be switched, as the striker's job is to kick the ball into the net and the goalkeeper's strategy involves diving to one side. Therefore, this penalty kick scenario is defined as a 'Non-symmetric Zero-sum' game.

\[
B=-A^{T}
\tag{2}
\]

To analyze game dynamics and how player strategies evolve, replicator dynamics was used to analyze this system. Replicator dynamics involves analyzing mixed-strategy games that evolve over continuous time and learning between agents over time. Since the game is zero-sum, the total payoff is zero [4], which means:

\[
x^{T} A y + y^{T} B x = 0
\tag{3}
\]

Since our game is a two-player game, the replicator dynamics can be defined as \cite{engel2023}:

\begin{align}
\dot{x}_{i} &= x_{i} \left( \{A y\}_{i} - x^{T} A y \right) \tag{4} \\
\dot{y}_{j} &= y_{j} \left( \{B x\}_{j} - y^{T} B x \right) \tag{5}
\end{align}

Equation (4) covers the replicator dynamics for the striker and how their strategies evolve. $\{A y\}_{i}$ is the payoff for the striker playing strategy $i$, given the goalkeeper plays a mixed strategy y and the replicator dynamics compares the payoffs of the pure strategy against the average payoff obtained using the mixed strategy $x$, which is denoted by $x^{T} A y$. Equation (5) denotes the same thing but from the goalkeeper's perspective, with $\{B x\}_{j}$ denoting the payoff for the goalkeeper's current strategy against the striker's strategy mix of x and $y^{T} B x$ being the average payoff for the goalkeeper. Therefore, these dynamics allow strategies that outperform the average payoff to be more prevalent and strategies that underperform to fade away. This allows the players to adapt their strategy to the other player as they aim to optimize their payoffs.
For this game, x and y are $3 \times 1$ column vectors which represent the probability distribution for the kicker's strategies and goalkeeper strategies, respectively, which are defined as:

\[
\begin{aligned}
& i \in \{\text{kickleft, kickcenter, kickright}\} = \{1,2,3\} \\
& j \in \{\text{diveleft, divecenter, diveright}\} = \{1,2,3\}
\end{aligned}
\]

Since the game is non-cooperative, there exists at least one Nash equilibrium, which is true for our system. Nash equilibrium for a system is defined as a point where unilaterally changing the strategy does not benefit \cite{fryer2018}. For replicator dynamics, a pair of mixed strategies $x^{*}$ and $y^{*}$ are the Nash equilibrium point if the agents interpret their respective payoffs to be optimized. For zero-sum games, Nash strategies are also referred to as max-min strategies. The following expressions describe the criteria for the Nash equilibrium \cite{engel2023}:

\begin{align}
\forall i \in [n]: & \quad (A y^*)_i = x^{*T} A y^* \tag{6} \\
\forall j \in [n]: & \quad (B x^*)_j = y^{*T} B x^* \tag{7}
\end{align}

By setting equations (4) and (5) for all the possible strategies and setting them to 0 such that $\frac{d x_{i}}{dt}=0$ and $\frac{d y_{j}}{dt}=0$, the equilibrium strategy mixes can be calculated and compared with equations (6) and (7), it can be determined whether or not the strategies obtained are Nash or not. Pure strategies can also be fixed points in equations (6) and (7); however, those will be exploited by the other player easily and therefore are not considered to be a Nash equilibrium for the game.

\section{Lyapunov Stability Analysis}

Since our penalty kick game is dynamic and nonlinear, a suitable Lyapunov candidate function can be used to analyze our two-agent system. From existing literature, the Kullback-Leibler (K-L) divergence can be used as a Lyapunov function as defined below \cite{allesina2025}:

\[
V(x, y) = \sum_{i=1}^{3} x_{i}^{*} \ln \left( \frac{x_{i}^{*}}{x_{i}} \right ) + \sum_{j=1}^{3} y_{j}^{*} \ln \left( \frac{y_{j}^{*}}{y_{j}} \right )
\tag{8}
\]

The stability criteria using Lyapunov function \cite{khalil2001}:

\begin{itemize}
\item $\frac{d V}{d t} \leq 0$: system is stable
\item $\frac{d V}{d t} < 0$: system is asymptotically stable
\item $\frac{d V}{d t} > 0$: system is unstable
\end{itemize}

The KL divergence measures how different the current strategy mix is different than the Nash equilibrium. It meets the following conditions:

\begin{itemize}
\item $V(x^*, y^*) = 0$ and this is the only zero point of the function
\item $V(x, y) > 0$ for all $x \neq x^{*}, y \neq y^{*}$
\end{itemize}

The time derivative equation (8) ends up being 0 at the Nash equilibrium, meaning that the Nash equilibrium is neutrally stable, such that if they started at the equilibrium, the system stays there, but starting at a local neighborhood of the equilibrium, the trajectories for the strategy will oscillate about the equilibrium point. The proof for this is shown below:\\

Proof 1: Neutral stability

\[
\begin{gathered}
\dot{V} = -\sum_{i=1}^{3} \frac{x_{i}^{*}\left(\dot{x_{i}}\right)}{x_{i}} - \sum_{j=1}^{3} \frac{y_{j}^{*}\left(\dot{y_{j}}\right)}{y_{j}} \\
\dot{x_{i}} = x_{i}\left(\{A y\}_{i}-x^{T} A y\right) \text{ and } \dot{y_{j}} = y_{j}\left(\{B x\}_{j}-y^{T} B x\right) \\
\dot{V} = - \left( \sum_{i=1}^{3} x_{i}^{*} \frac{(x_{i}\left(\{A y\}_{i}-x^{T} A y\right))}{x_{i}} + \sum_{j=1}^{3} y_{j}^{*} \frac{(y_{j}\left(\{B x\}_{j}-y^{T} B x\right))}{y_{j}} \right ) \\
\dot{V} = - \left( \sum_{i=1}^{3} x_{i}^{*}\left(\{A y\}_{i}-x^{T} A y\right) + \sum_{j=1}^{3} y_{j}^{*}\left(\{B x\}_{j}-y^{T} B x\right) \right ) x^{T}A y + y^{T}B x = 0 \\
\dot{V} = - \left( \sum_{i=1}^{3} x_{i}^{*}\left(\{A y\}_{i}\right) + \sum_{j=1}^{3} y_{j}^{*}\left(\{B x\}_{j}\right) \right ) B = -A^{T} \\
\dot{V} = - (x^{*} - x)^{T}A(y - y^{*}) \text{ For Nash equilibrium } (x, y) = (x^{*}, y^{*}),\ \dot{V} = 0 \text{ neutrally stable }
\end{gathered}
\]

Near equilibrium: $\dot{V} = \delta x^{T} A \delta y \rightarrow \delta x \approx 0$ and $\delta y \approx 0$, then $\dot{V} \approx 0$\\

This means if the system starts at the equilibrium, it stays there; otherwise, it oscillates around the equilibrium without ever converging to it if the initial conditions are close enough to the equilibrium \cite{khalil2001}. Since the payoff matrix A is not symmetric (i.e. $A \neq A^{T}$ ) but is a real matrix, then the matrix has to be averaged to make it symmetric to analyze definiteness of the real matrix A using Johnson's method as shown below \cite{weisstein}:

\[
A_{s} = \frac{1}{2}(A + A^{T})
\]

This results in the following updated matrix $A_{s}$:

\[
A_s = 
\begin{bmatrix}
0 & 0.85 & 0.725 \\
0.85 & 0 & 0.325 \\
0.725 & 0.325 & 0
\end{bmatrix}
\]

This matrix is symmetric, so by finding the eigenvalues, the matrix definiteness can be determined. The eigenvalues are $\lambda_{1}=1.29, \lambda_{2}=-0.970, \lambda_{3}=-0.320$. Since the eigenvalues are both positive and negative, the matrix is determined to be indefinite and does not allow for inference of the behavior of the system for trajectories near the equilibrium. If the initial trajectories stay close to equilibrium, meaning $\delta x \approx 0$ and $\delta y \approx 0$, then $\dot{V} \approx 0$, resulting in the system being neutrally stable.
This means that if the system starts at equilibrium, the strategy mixes do not converge to the Nash equilibrium; rather, the game evolves into a cycle of outwitting the opponent and the opponent catching up, resulting in cyclical dynamics for the system.

\section{Simulation and results}

Using the background and game setup, this problem was simulated in Python (code attached to this project report). The Nash equilibria for the system were as follows:

\[
x^{*} = [0.503,0.422,0.075], y^{*} = [0.416,0.276,0.308]
\]

This shows that the striker prefers to shoot left and center more often than to the right, while the goalkeeper adopts a more balanced approach as to which direction to dive towards, with the left being preferable since the striker tends to shoot that way more often. Examining this equilibrium point using equations (6) and (7) proves that the system is a Nash equilibrium. Tables 1 and 2 show the results of this analysis for the striker and goalkeeper, respectively.

\begin{table}[h]
\centering
\begin{tabular}{cccc}
\toprule
Strategy & $\{A y\}_{i}$ & Optimal payoff for the striker $x^{*T}A y^{*}$ & $\Delta=\{A y\}_{i}-x^{*T}A y^{*}$ \\
\midrule
Kick left ($i=1$)   & 0.436 & 0.436 & 0 \\
Kick center ($i=2$) & 0.436 & 0.436 & 0 \\
Kick right ($i=3$)  & 0.436 & 0.436 & 0 \\
\bottomrule
\end{tabular}
\caption{Striker Payoff Analysis}
\end{table}

\begin{table}[h]
\centering
\begin{tabular}{cccc}
\toprule
Strategy & $\{B x\}_{j}$ & Optimal payoff for the goalkeeper $y^{*T}B x^{*}$ & $\Delta=\{B x\}_{j}-y^{*T}B x^{*}$ \\
\midrule
Dive left ($j=1$)   & -0.436 & -0.436 & 0 \\
Dive center ($j=2$) & -0.436 & -0.436 & 0 \\
Dive right ($j=3$)  & -0.436 & -0.436 & 0 \\
\bottomrule
\end{tabular}
\caption{Goalkeeper Payoff Analysis}
\end{table}

Starting the trajectories exactly at the equilibrium point, they stay there as shown in Figure 1:

\begin{figure}[h!]
  \centering
  \fbox{%
    \begin{minipage}{0.8\textwidth}
      \centering
      \includegraphics[width=\linewidth]{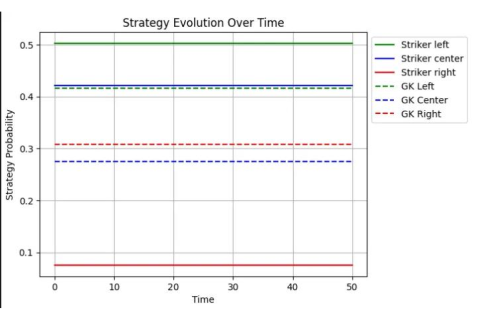}
    \end{minipage}
  }
  \caption{Initial conditions are the same as the Nash equilibrium, allowing the trajectories to stay where they started}
  \label{fig:init-cond}
\end{figure}

Lyapunov stability theory indicated that the system was neutrally stable, meaning that if we started in a neighborhood of the equilibrium point, the trajectories oscillate about the equilibrium point. Figure 2 shows this phenomenon happening for a randomly chosen initial probability distribution for the striker and goalkeeper.

\begin{figure}[h!]
  \centering
  \fbox{%
    \begin{minipage}{0.8\textwidth}
      \centering
      \includegraphics[width=\linewidth]{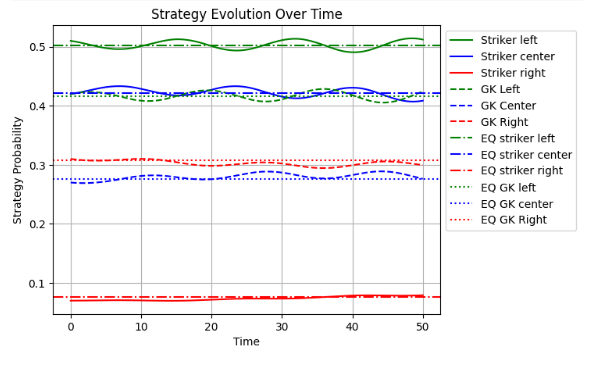}
    \end{minipage}
  }
  \caption{Strategy evolution over time for the initial probability distributions, x= [0.51,0.42,0.07] and y= [0.42,0.27,0.31]}
  
  \label{fig:init-cond}
\end{figure}

Further examining the system under varying perturbations, randomly generated Gaussian noise with a mean of 0 and varying standard deviations was added to the known equilibrium point, allowing for randomly generated initial conditions a neighborhood away from the equilibrium. Figures 3-4 show the generated trajectories with $\sigma=0.02$ and $\sigma=0.2$, respectively.

\begin{figure}[h!]
  \centering
  \fbox{%
    \begin{minipage}{0.8\textwidth}
      \centering
      \includegraphics[width=\linewidth]{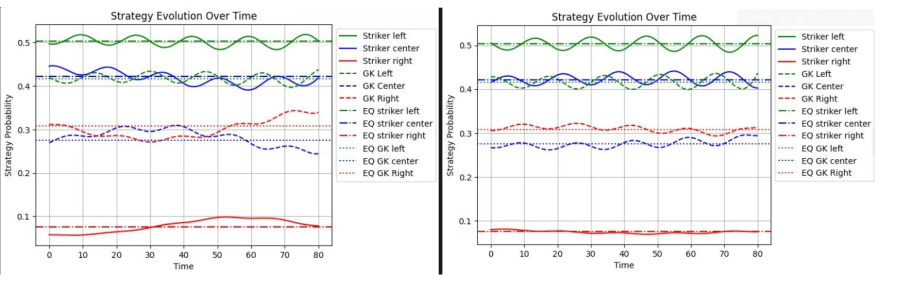}
    \end{minipage}
  }
\caption{(a) (b): Strategy evolution over time for the initial probability distributions, $\sigma=0.02$}
\end{figure}

\begin{figure}[h!]
  \centering
  \fbox{%
    \begin{minipage}{0.8\textwidth}
      \centering
      \includegraphics[width=\linewidth]{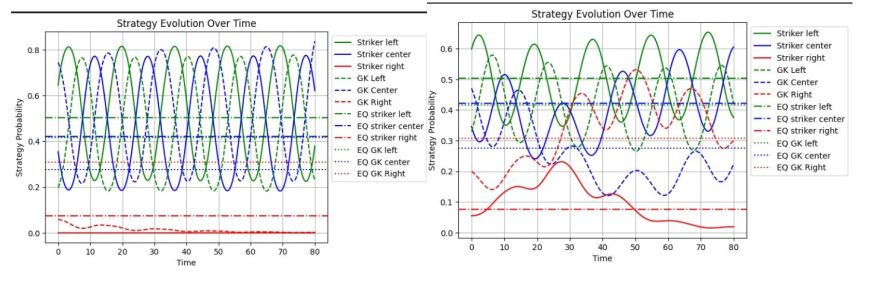}
    \end{minipage}
  }
\caption{(a) (b): Strategy evolution over time for the initial probability distributions, $\sigma=0.2$}
\end{figure}

It was observed that for these initial conditions, the trajectories kept oscillating within a neighborhood of the equilibrium, meaning that neither agent learned the optimal strategy for this penalty shootout, leaving more to be desired to achieve a strategic balance between the striker and the goalkeeper. Observing from Figures 3 and 4, increasing the width of the distribution of the noise increases the distance between the initial conditions and the equilibrium point of the system. This results in an increase in oscillations around the equilibrium point and reflects the neutral stability of the system as highlighted by the Lyapunov theory. To examine this further, the payoffs for each player over time are analyzed. Modifying equations (6) and (7) allows us to examine how the player strategies are paying off for the striker and goalkeeper concerning the average payoff for each player:

\[
\{A y\}_{i}-x^{T} A y \neq 0 \text{ and }\{B x\}_{j}-y^{T} B x \neq 0
\]

Comparing the payoffs against the average for both players through plotting these differences for figures 3(a) and 4(a), we obtain the following plots as shown in figures 5-6:

\begin{figure}[h!]
  \centering
  \fbox{%
    \begin{minipage}{0.8\textwidth}
      \centering
      \includegraphics[width=\linewidth]{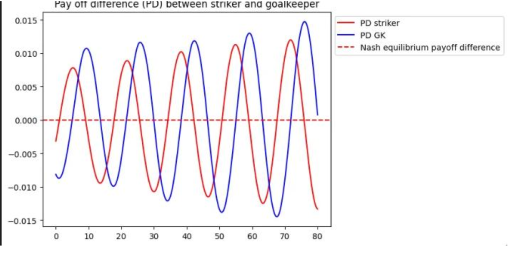}
    \end{minipage}
  }
\caption{Payoff difference for 3(a) against time (seconds)}
\end{figure}

\begin{figure}[h!]
  \centering
  \fbox{%
    \begin{minipage}{0.8\textwidth}
      \centering
      \includegraphics[width=\linewidth]{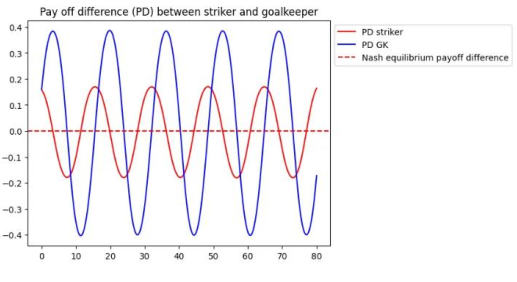}
    \end{minipage}
  }
\caption{Payoff difference for 4(a) against time(seconds)}
\end{figure}

The payoff differences for both the striker and goalkeeper oscillate about 0, meaning that neither player was able to reach the Nash equilibrium, and the players were constantly adapting to each other's strategy. The increased width of the Gaussian distribution allowed for the differences to be more pronounced, meaning that the trajectory for the payoffs was deviating further away from the equilibrium. A positive difference indicates that the current strategy is generating a higher payoff than the average for a player, and vice versa, meaning the strategies that perform better are being chosen while underperforming ones are diminishing. Both the strikers' and the goalkeepers' payoff differences also move in the same direction, indicating that they are adapting to their opponent's strategy.
This results in the game evolving into a cycling of adapting to each other's strategies, and since the strategies do not converge to the Nash equilibrium, it leaves players with incentives to be able to benefit from unilaterally changing their strategies. Since the strategic balance is never achieved by the replicator dynamics themselves, more work had to be done through feedback linearization to achieve the strategic balance, allowing for more predictability in the game.

\section{Feedback Linearization}

Since replicator dynamics by themselves proved to be ineffective in teaching the agents the optimal strategy, full-state feedback linearization was introduced to the system. Feedback linearization allows the system to be globally asymptotically stable \cite{khalil2001}, allowing convergence to the Nash equilibrium regardless of the initial starting condition. This allows both agents to adapt the optimized strategy for both agents so that they maximize their payoffs relative to the other player. Equations (9) and (10) show the updated replicator dynamics for feedback linearization being added to the system.

\begin{align}
\frac{d x_{i}}{d t}=x_{i}\left(\{A y\}_{i}-x^{T} A y\right)+u_{i}, i \in\{1,2,3\} \tag{9} \\
\frac{d y_{j}}{d t}=y_{j}\left(\{B x\}_{j}-y^{T} B x\right)+v_{j}, j \in\{1,2,3\} \tag{10}
\end{align}

Since each of the systems has a degree of 3, due to three state equations, there must be 3 control outputs for each agent, meaning a total of 6 control outputs to the system. The full state linearization can be defined by the following equation \cite{khalil2001}:

\[
\begin{gathered}
\frac{d x}{d t}=f(x)+g(x) u \\
y=h(x) \tag{11}
\end{gathered}
\]

For the updated dynamics as shown in equations (9) and (10), equation (11) parameters for the striker and goalkeeper can be defined as:

\begin{align}
& \frac{d x}{d t}=
\begin{bmatrix}
\frac{d x_{1}}{d t} \\
\frac{d x_{2}}{d t} \\
\frac{d x_{3}}{d t}
\end{bmatrix},
f(x)=
\begin{bmatrix}
x_{1}\{A y\}_{1}-x^{T} A y \\
x_{2}\{A y\}_{2}-x^{T} A y \\
x_{3}\{A y\}_{3}-x^{T} A y
\end{bmatrix},
g(x)=
\begin{bmatrix}
1 & 0 & 0 \\
0 & 1 & 0 \\
0 & 0 & 1
\end{bmatrix},
u=
\begin{bmatrix}
u_{1} \\
u_{2} \\
u_{3}
\end{bmatrix}
\tag{12} \\
& \frac{d y}{d t}=
\begin{bmatrix}
\frac{d y_{1}}{d t} \\
\frac{d y_{2}}{d t} \\
\frac{d y_{3}}{d t}
\end{bmatrix},
f(y)=
\begin{bmatrix}
y_{1}\{B x\}_{1}-y^{T} B x \\
y_{2}\{B x\}_{2}-y^{T} B x \\
y_{3}\{B x\}_{3}-y^{T} B x
\end{bmatrix},
g(y)=
\begin{bmatrix}
1 & 0 & 0 \\
0 & 1 & 0 \\
0 & 0 & 1
\end{bmatrix},
v=
\begin{bmatrix}
v_{1} \\
v_{2} \\
v_{3}
\end{bmatrix}
\tag{13}
\end{align}

Since the dynamics of the system are well-known in this instance, it is convenient to cancel out the non-linearities in the system and add a linear term that involves the convergence of the probability for each strategy to reach the Nash equilibrium. The control inputs $u$ and $v$ are the following:

\[
u=
\begin{bmatrix}
-x_{1}\{A y\}_{1}+x^{T} A y-k_{1}\left(x_{1}-x_{1}^{*}\right) \\
-x_{2}\{A y\}_{2}+x^{T} A y-k_{2}\left(x_{2}-x_{2}^{*}\right) \\
-x_{3}\{A y\}_{3}+x^{T} A y-k_{3}\left(x_{3}-x_{3}^{*}\right)
\end{bmatrix}
\]
\[
v=
\begin{bmatrix}
-y_{1}\{B x\}_{1}+y^{T} B x-c_{1}\left(y_{1}-y_{1}^{*}\right) \\
-y_{2}\{B x\}_{2}+y^{T} B x-c_{2}\left(y_{2}-y_{2}^{*}\right) \\
-y_{3}\{B x\}_{3}+y^{T} B x-c_{3}\left(y_{3}-y_{3}^{*}\right)
\end{bmatrix}
\tag{14}
\]

Inserting the control inputs $u$ and $v$ displayed in equation (14) into equations (12) and (13), the nonlinearities of the replicator dynamics are cancelled out, which results in the following dynamic systems equations:

\[
\begin{bmatrix}
\frac{d x_{1}}{d t} \\
\frac{d x_{2}}{d t} \\
\frac{d x_{3}}{d t}
\end{bmatrix}
=
\begin{bmatrix}
-k_{1}\left(x_{1}-x_{1}^{*}\right) \\
-k_{2}\left(x_{2}-x_{2}^{*}\right) \\
-k_{3}\left(x_{3}-x_{3}^{*}\right)
\end{bmatrix}
,
\begin{bmatrix}
\frac{d y_{1}}{d t} \\
\frac{d y_{2}}{d t} \\
\frac{d y_{3}}{d t}
\end{bmatrix}
=
\begin{bmatrix}
-c_{1}\left(y_{1}-y_{1}^{*}\right) \\
-c_{2}\left(y_{2}-y_{2}^{*}\right) \\
-c_{3}\left(y_{3}-y_{3}^{*}\right)
\end{bmatrix},
\quad k_{i}, c_{i} > 0
\tag{15}
\]

Using equation (15), asymptotic stability for the system can be achieved for any value greater than zero to achieve global asymptotic stability. To prove this notion, $\frac{d x}{d t}$ was calculated again, and it was less than zero, proving the asymptotic stability as shown in Proof 2. The value of these constants determines the rate of convergence to the Nash equilibrium, with a higher value allowing for faster convergence and vice versa, and can be chosen arbitrarily if they are both greater than zero. \\

Proof 2: Asymptotic stability

\[
\begin{gathered}
\dot{V}=-\sum_{i=1}^{3} \frac{x_{i}^{*}\left(\dot{x_{i}}\right)}{x_{i}}-\sum_{j=1}^{3} \frac{y_{j}^{*}\left(\dot{y_{j}}\right)}{y_{j}} \\
\dot{x_{i}}=-k_{i}\left(x_{i}-x_{i}^{*}\right), \dot{y_{j}}=-c_{j}\left(y_{j}-y_{j}^{*}\right) \\
\dot{V}=-(\sum_{i=1}^{3} \frac{x_{i}^{*}\left(-k_{i}\left(x_{i}-x_{i}^{*}\right)\right)}{x_{i}}+\sum_{j=1}^{3} \frac{y_{j}^{*}\left(-c_{j}\left(y_{j}-y_{j}^{*}\right)\right)}{y_{j}}) \\
\dot{V}=-(\sum_{i=1}^{3} x_{i}^{*}\left(k_{i}\left(-\frac{x_{i}}{x_{i}}+\frac{x_{i}^{*}}{x_{i}}\right)\right)+\sum_{j=1}^{3} y_{j}^{*}\left(c_{j}\left(-\frac{y_{j}}{y_{j}}+\frac{y_{j}^{*}}{y_{j}}\right)\right)
\end{gathered}
\]

$\dot{V}=\sum_{i=1}^{3} x_{i}^{*}\left(k_{i}\left(1-\frac{x_{i}^{*}}{x_{i}}\right)\right)+\sum_{j=1}^{3} y_{j}^{*}\left(c_{j}\left(1-\frac{y_{j}^{*}}{y_{j}}\right)\right)$, for $\frac{y_{j}^{*}}{y_{j}}, \frac{x_{i}^{*}}{x_{i}}>1 \dot{V}<0$

However, this fails to prove the global asymptotic stability proposed by the feedback linearization theorem due to the constraints, meaning that if the probability for a given direction is greater than 1, the proof for global asymptotic stability breaks down. 
Therefore, a different Lyapunov function must be used to prove global asymptotic stability. Consider:

\[
V(x, y)=\sum_{i=1}^{3} \frac{1}{2}\left(x_{i}-x_{i}^{*}\right)^{2}+\sum_{j=1}^{3} \frac{1}{2}\left(y_{j}-y_{j}^{*}\right)^{2}
\]

This candidate function also meets the following conditions:

\begin{itemize}
\item $V\left(x^{*}, y^{*}\right)=0$ and this is the only zero point of the function
\item $V(x, y)>0$ for all $x \neq x^{*}, y \neq y^{*}$
\end{itemize}

Proof 2 with the new Lyapunov candidate:

\[
\begin{gathered}
\dot{V}(x, y)=\sum_{i=1}^{3} \dot{x_{i}}\left(x_{i}-x_{i}^{*}\right)+\sum_{j=1}^{3} \dot{y_{j}}\left(y_{j}-y_{j}^{*}\right) \\
\dot{x_{i}}=-k_{i}\left(x_{i}-x_{i}^{*}\right), \dot{y_{j}}=-c_{j}\left(y_{j}-y_{j}^{*}\right) \\
\dot{V}(x, y)=\sum_{i=1}^{3}-k_{i}\left(x_{i}-x_{i}^{*}\right)\left(x_{i}-x_{i}^{*}\right)-\sum_{j=1}^{3} c_{j}\left(y_{j}-y_{j}^{*}\right)\left(y_{j}-y_{j}^{*}\right) \\
\dot{V}(x, y)=\sum_{i=1}^{3}-k_{i}\left(x_{i}-x_{i}^{*}\right)^{2}+\sum_{j=1}^{3}-c_{j}\left(y_{j}-y_{j}^{*}\right)^{2},\ \forall k, c>0 \dot{V}<0 \rightarrow \text {asymptotic stability}
\end{gathered}
\]

This second candidate Lyapunov function was able to prove that the dynamics are globally asymptotically stable. This means that regardless of the starting point, the dynamics will guarantee convergence to the Nash equilibrium, resulting in both players learning the optimal strategy for the penalty shootout.

\begin{figure}[h!]
  \centering
  \fbox{%
    \begin{minipage}{0.8\textwidth}
      \centering
      \includegraphics[width=\linewidth]{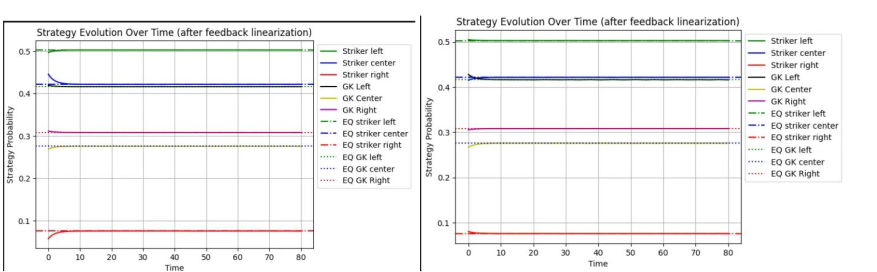}
    \end{minipage}
  }
\caption{(a) (b): Strategy evolution over time after feedback linearization, $\sigma=0.02, k = c = 0.5 $}
\end{figure}

\begin{figure}[h!]
  \centering
  \fbox{%
    \begin{minipage}{0.8\textwidth}
      \centering
      \includegraphics[width=\linewidth]{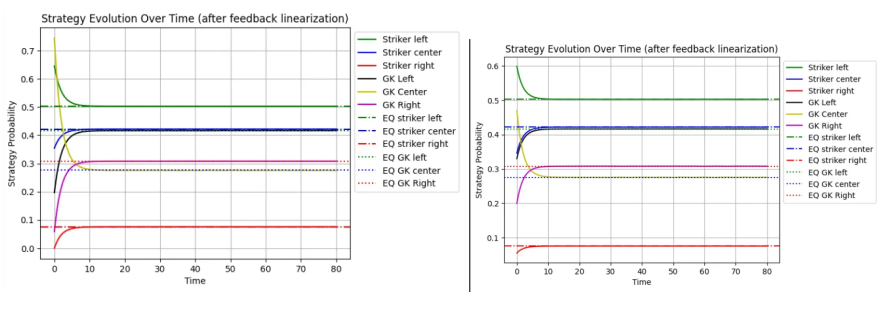}
    \end{minipage}
  }
\caption{(a) (b): Strategy evolution over time after feedback linearization, $\sigma=0.2, k = c = 0.5$}
\end{figure}

Figures 7 and 8 display that regardless of the magnitude of the perturbation added to the initial conditions, the feedback linearization controls displayed in equation (15) allow the player strategies to converge to the desired Nash equilibrium of the system, making the system asymptotically stable. To further visualize the asymptotic stability of the system, the payoff deviation from the average can be analyzed since the convergence to the Nash condition suggests that the payoff difference (PD) should converge to 0, signaling the convergence to the Nash equilibrium. To illustrate this concept, figures 9 and 10 are displayed below with both the standard deviation values.

Compared to the neutral stability of the replicator dynamics by themselves, the feedback linearization guarantees convergence to the Nash equilibrium, meaning that players choose a strategy that does as well as the optimal mixed strategy. Therefore, by converging to the Nash equilibrium, this feedback linearization allows both players to learn their optimal strategies and achieve balance in the game to prevent it from cycling forever and minimizes the risk of exploitation or bugs in the game, creating a strategic balance between the player and the goalkeeper. In real life, this can be imagined as coaching based on scouting reports and data from previous penalty shootouts, which allow players to optimize their respective strategies in a penalty shootout. The system stabilizes- not because everyone is winning, but because neither player wins by unilaterally changing their strategy.

\begin{figure}[h!]
  \centering
  \begin{minipage}{0.49\linewidth}
    \includegraphics[width=\linewidth]{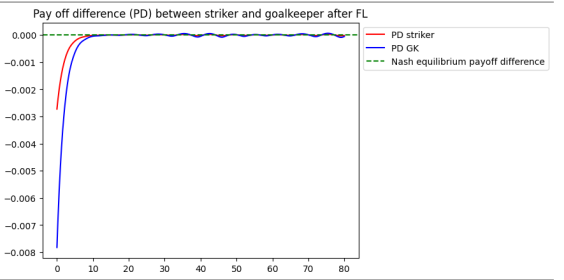}
    \caption{PD after FL was implemented 7(a)}
  \end{minipage}
  \begin{minipage}{0.49\linewidth}
    \includegraphics[width=\linewidth]{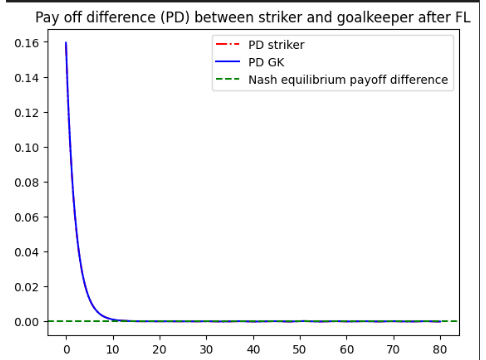}
    \caption{PD after FL was implemented 8(a)}
  \end{minipage}
\end{figure}

\section{Limitations and Future Work}

This framework relies on a number of large assumptions that limit the scope of immediate practical application. First, the framework requires exact knowledge of the payoff matrix $A$, usually unavailable in real-world strategic interactions, as payoffs must come from noisy observations or a moving target that alters the game conditions. Second, the implementation assumes centralized control capabilities, such that an external coordinator could modify players' strategy evolution rates within each population---a strong assumption that may not indicate the autonomous orientation of decision-makers who experience strategy evolution. Third, the analysis focuses on the interior of the probability simplex, and more consideration will be necessary to capture behavior close to the boundary of strategies ($x_{i}=0 $) by projection methods or via barrier functions. In the numerical implementation, clipping and normalization techniques were used to maintain the probability constraints of sums not exceeding 1 and probability values being always positive.  Finally, the approach assumes players have continuous strategy adjustments that follow the evolutionary differential equations; however, strategic behavior as we know it normally consists of discrete decisions that have inherent cognitive or physical limitations.

These constraints suggest several useful extensions. First, which would limit the robustness to convergence to the Nash equilibrium concerning model uncertainty through models of uncertainty, the motivation of sliding mode control is that it would limit the error between the actual and estimated systems. Second, the assumption of fixed payoff matrices could be loosened to allow for time-varying games and player-specific attributes, which allow for more realistic strategic modeling. Third, distributed implementation methods could alleviate the need for centralized control, such that each autonomous player can converge to Nash with limited knowledge based on local interactions. Finally, discrete-time formulations could be constructed to show more realistic transitions between the continuous dynamics theoretically and the discrete decision-making process. 

\section{Conclusion:}

In this paper, we demonstrate the first use of feedback linearization to simplify the replicator dynamics framework, which provides a framework for achieving guaranteed convergence in evolutionary games from a control-theoretic perspective. Rigorous Lyapunov analysis was able to demonstrate that the control law produces global asymptotic stability, which modifies the oscillatory natural behavior of replicator dynamics into a predictable convergence to Nash equilibria. 

We showed that this methodology illustrates how control theory can fundamentally alter the convergence properties of evolutionary games, which traditional learning approaches do not create theoretical guarantees for, and while we illustrated these properties through the example of the penalty kick analysis, the theory would apply to any type of strategic interaction described by replicator dynamics.

Future research will unfold this foundational work using sliding mode control to accommodate unknown payoff structures, distributed implementations acting as autonomous agents, and a discrete-state formulation in real-world settings. Establishing this new methodological approach creates a new stream of research from the intersection of control theory and evolutionary games, which can now be utilized in several fields such as multi-agent systems, algorithmic trading, and strategic optimization.

\section*{Appendix:}
GitHub repo for this project:\\
\url{https://github.com/adilfaisal01/SE762--Game-theory-and-Lyapunov-calculations}

Feel free to leave comments and questions, and I will be happy to answer them.

\end{document}